\documentclass[journal=jacsat,manuscript=article]{achemso}

\usepackage[version=3]{mhchem} % Formula subscripts using \ce{}
\usepackage[T1]{fontenc}       % Use modern font encodings

\author{Jose H. Garcia}
\email{ josehugo.garcia@icn2.cat}
\author{Aron W. Cummings}
\author{Stephan Roche}
\affiliation[Main University]
{Catalan Institute of Nanoscience and Nanotechnology (ICN2), CSIC and The Barcelona Institute of Science and Technology,  Campus UAB, 08193 Barcelona, Spain. }
\alsoaffiliation[Second University]
{ICREA - Instituci\'o Catalana de Recerca i Estudis Avan\c{c}ats, 08010 Barcelona, Spain. }

\title[An \textsf{achemso} demo]
  { Spin Hall effect and Weak Antilocalization in Graphene/Transition Metal Dichalcogenide Heterostructures}

\abbreviations{IR,NMR,UV}

\usepackage{bm}
\usepackage{color}
 \usepackage{soul} % para tachar palabras
\renewcommand{\d}[2]{\frac{d #1}{d #2}} % for derivatives
\newcommand{\medio}[1]{\left\langle #1 \right\rangle}

\begin{document}

\begin{tocentry}

Some journals require a graphical entry for the Table of Contents.
This should be laid out ``print ready'' so that the sizing of the
text is correct.

Inside the \texttt{tocentry} environment, the font used is Helvetica
8\,pt, as required by \emph{Journal of the American Chemical
Society}.

The surrounding frame is 9\,cm by 3.5\,cm, which is the maximum
permitted for  \emph{Journal of the American Chemical Society}
graphical table of content entries. The box will not resize if the
content is too big: instead it will overflow the edge of the box.

This box and the associated title will always be printed on a
separate page at the end of the document.

\end{tocentry}

\begin{abstract}
We report on a theoretical study of the spin Hall Effect (SHE) and weak antilocalization (WAL) in graphene/transition metal dichalcogenide (TMDC) heterostructures, computed through efficient real-space quantum transport methods, and using realistic tight-binding models parametrized from {\it ab initio} calculations. The graphene/WS$_2$ system is found to maximize spin proximity effects compared to graphene on MoS$_2$, WSe$_2$, or MoSe$_2$,  with a crucial role played by disorder, given the disappearance of SHE signals in the presence of strong intervalley scattering. Notably, we found that stronger WAL effects are concomitant with weaker charge-to-spin conversion efficiency. For further experimental studies of graphene/TMDC heterostructures, our findings provide guidelines for reaching the upper limit of spin current formation and for fully harvesting the potential of two-dimensional materials for spintronic applications.
\end{abstract}

\section{Introduction}

Van der Waals heterostructures, made by assembling different classes of two-dimensional materials, represent a fascinating playground for materials innovation \cite{Geim2014,Miro2014,Novoselov2016}. The improvement of the material growth techniques is giving rise to intense efforts in developing practical applications such as non-volatile resistive memory devices \cite{Hua1}, energy harvesting \cite{Ferrari2015}, or optical detectors \cite{Liangbing2016}. Among the variety of promising uses of these heterostructures, spintronics stands as a leading topic which could shortly witness the full practical realization of innovative devices \cite{Roche2015, Kawakami2015}. Indeed, beyond the long room-temperature spin lifetime achieved experimentally \cite{Drogeler2014,Drogeler2016,Ingla2016}, evidences of spin transport modulation by proximity effects have been reported in heterostructures combining graphene with magnetic insulators \cite{Leutenantsmeyer2016, Yang2013, Hallal2017, Singh2017}, or by enhancing spin-orbit coupling (SOC) in graphene through the deposition of adatoms \cite{Avsar2015, Cresti2016, VanTuan2016}. Another route is the deposition of graphene onto transition metal dichalcogenides (TMDCs) \cite{Schmidt2016, Ge2017}, where a considerable enhancement of SOC in graphene interfaced with $\text{WS}_2$ or $\text{WSe}_2$ has been unveiled by measuring weak antilocalization (WAL) effects \cite{Wang2015, Wang2016, Yang2016}.

Additionally, large spin Hall effect signals have been reported in graphene/WS$_2$ heterostructures \cite{Avsar2014}, suggesting that such interfaces could be suitable for the optimization of pure spin currents and the development of spin-torque technologies \cite{Roche2015}. However, the observation of SHE in graphene/TMDC systems has been difficult to reproduce, opening some debate about how intrinsic mechanisms inducing Hall effects are actually related to non-local resistance signals \cite{Kaverzin2015, Niimi2015, Wang2015, Wang2015b, VanTuan2016, Cresti2016}. Additionally, some recent {\it ab initio} studies of the electronic structure of graphene/TMDC layers have shown that the inherited SOC parameters in graphene lie in the meV range \cite{Gmitra2015}, and little is known about how such SOC proximity effects impact spin dynamics and the emergence of SHE. Similarly, to date the origin of the measured large WAL effects in graphene/TMDC heterostructures, as well as the possibility to simultaneously observe the formation of the SHE, are unclear and challenging, and demand a detailed inspection of the joint contribution of disorder and SOC proximity effects.

Here, we use a realistic tight-binding model, parametrized from {\it ab initio} simulations, to exactly compute spin Hall conductivities and dissipative conductivities in graphene/TMDC heterostructures using Kubo transport methods. We show that a sizable SHE signal should be observable in such heterostructures, and by varying the disorder strength, we establish a range of experimental conditions for its optimization. The largest SHE signal is obtained for graphene on WS$_2$ when disorder-induced scattering is limited to intravalley processes. In contrast, the presence of strong intervalley scattering, essential for observing WAL effects, leads to a large reduction of the SHE figure of merit. This shows that although WAL is an extremely useful tool to probe the efficiency of proximity-induced SOC in graphene \cite{Wang2015, Wang2016, Yang2016, Grbic2008, Newton2017}, its presence precludes the existence of a SHE signal that is large enough for practical use. Therefore, to maximize the magnitude of the intrinsic SHE in graphene/TMDC heterostructures, the highest interface quality should be sought.

\section{Model and Methods}
 
The electronic properties of graphene on a TMDC substrate can be captured with the Hamiltonian
$H=H_{\text{orb}}+H_{\text{so}}$. The first term \cite{Gmitra2016, Kochan2017} ,
\begin{equation}
H_{\text{orb}}= -t\sum_{\medio{i,j}} \left(a_i^\dagger b_j+ b_i^\dagger a_i\right)+ \frac{\Delta}{2}\sum_{i} (a_i^\dagger a_i-b_i^\dagger b_i),
\end{equation}
represents the orbital part of the Hamiltonian, where the operators $a_i^\dagger\,(b_i^\dagger)$ and $a_i\,(b_i)$ create and annihilate respectively an electron on site $i$ of graphene's $A(B)$ sublattice. The nearest-neighbor hopping is given by $t$, while $\Delta$ accounts for a weak superlattice effect induced by the TMDC, which creates a gap of magnitude $\Delta$ and modifies the Fermi velocity of graphene. The second term,
\begin{align}
H_{\text{so}}&=\frac{2i}{3}\sum_{\medio{i,j},\sigma}(\hat{\bm{s}}\times \bm{d}_{i,j})_{z,\sigma,\bar{\sigma}}~ \lambda_{\text{R}}\, a_{i,\sigma}^\dagger\, b_{j,\bar{\sigma}} +h. c \nonumber\\
&+\frac{2i}{3}\sum_{\medio{\medio{i,j}},\sigma}(\hat{\bm{s}}\times \bm{D}_{i,j})_{z,\sigma,\bar{\sigma}}\left( \lambda_{\text{PIA}} ^{(A)}\, a_{i,\sigma}^\dagger a_{j,\bar{\sigma}} + \lambda_{\text{PIA}} ^{(B)}\, b_{i,\sigma}^\dagger b_{j,\bar{\sigma}} \right)\nonumber\\
&+\frac{i}{3\sqrt{3}}\sum_{\medio{\medio{i,j}},\sigma } \nu_{i,j} (\hat{s}_z)_{\sigma,\sigma}\, ( \lambda_{I}^{(A)} a_{i,\sigma}^\dagger a_{j,\sigma}-\lambda_{I}^{(B)} b_{i,\sigma}^\dagger b_{j,\sigma}),
\end{align}
represents the proximity-induced enhancement of SOC in graphene due to a weak hybridization with the heavy atoms in the TMDC. It is composed of different types of SOC allowed by the symmetries of the system \cite{Gmitra2016, Kochan2017}. The first element is a Rashba SOC with strength $\lambda_{\text{R}}$ due to the broken out-of-plane symmetry, where $\bm{d}_{i,j}$ is a normalized vector pointing from site $i$ to its nearest neighbor at site $j$, and $\bm{s}$ is the normalized spin operator. The second is a sublattice-dependent PIA SOC \cite{Gmitra2016,Kochan2017}, where $\lambda_{\text{PIA}} ^{(A)}$ and $\lambda_{\text{PIA}} ^{(B)}$ are the coupling intensities in each sublattice and $\bm{D}_{i,j}$ is a normalized vector pointing from site $i$ to its next nearest neighbor at site $j$. The final term is a sublattice-dependent intrinsic SOC, with couplings intensities $\lambda_{\text{I}}^{(A)}$ and $\lambda_{\text{I}}^{(B)}$ for each sublattice. We use parameters provided in reference\cite{Gmitra2016}, which were extracted from fitting to DFT band structures and spin textures.

The disorder is incorporated by considering a random distribution of $n_\text{p}$ electron-hole puddles defined by the potential \cite{Adam2011,Ortmann2011}
\begin{equation}
U_n(\bm{r})= u_n \text{exp}\left(- \frac{(\bm{r}-\bm{R}_n)^2}{2\xi_\text{p}^2}\right),
\end{equation}
where $u_n$ is the puddle height randomly chosen within the range $[-U_\text{p}, U_\text{p}]$, $\bm{R}_n$ is the position of the center of the Gaussian puddle, and $\xi_\text{p}$ is the puddle range, fixed in our simulations as $\xi_\text{p}=\sqrt{3} a$ with $a$ being the lattice constant. Previous characterization of this type of 
puddle showed that one can induce a transition from intra- to intervalley-driven scattering in graphene by fine-tuning its parameters \cite{Zhang2009, Roche2012}.

The zero-temperature dissipative DC conductivity $\sigma_{\text{xx}}$ is calculated using a real-space $\mathcal{O}(N)$ numerical implementation of the Kubo-Greenwood formula based on wavepacket propagation \cite{Roche1997, Roche1999, L.E.F.FoaTorresS.Roche2014}. In this approach, the time-dependent diffusion coefficient at a given Fermi energy $D(\varepsilon_F,t)$ is obtained as the time derivative of the mean square displacement of the wavepacket $\Delta X^2 (\varepsilon_F,t) = |\langle X(t) \rangle^2- \langle X(t)^2 \rangle|$, i.e.,
\begin{equation}
D(\varepsilon_F,t)=\frac{\partial}{\partial t} \Delta X^2(\varepsilon_F,t),
\end{equation}
and is then related to the conductivity through the Einstein relation
\begin{equation}
\sigma_{\text{xx}}(\varepsilon_F)=\lim_{t\rightarrow \infty }\,\frac{1}{2}e^2\rho(\varepsilon_F)\, D(\varepsilon_F,t)
\end{equation}
where $\rho(\varepsilon_F)=\text{Tr}\left[\delta(H-\varepsilon_F)\right]$ is the density of states and $e$ is the electron charge. A computational advantage of this approach is that it can efficiently access all possible conduction mechanisms, from ballistic to diffusive and localized regimes \cite{Ortmann2011,Roche2012,Fan2014, Uppstu2014}. The diffusive conductivity occurs when the diffusion coefficient reaches a saturation limit (maximum value), where $\sigma_{\text{sc}}(\varepsilon_F)=\,\frac{1}{2}e^2\rho(\varepsilon_F) D_{\text{max}}(\varepsilon_F,t)$ \cite{L.E.F.FoaTorresS.Roche2014}, while at longer times the contribution of quantum corrections $\delta\sigma$, due to multiple scattering and interference effects, are encoded into the scaling behavior of the quantum conductivity $ \sigma_{\text{xx}}(\varepsilon_F)=\sigma_{\text{sc}}(\varepsilon_F)+\delta\sigma(\varepsilon_F)$. In the presence of weak disorder, the quantum correction is dictated by the scaling theory of localization \cite{Lee1985,Mucciolo2010} and takes the form
\begin{equation}
\delta\sigma=\pm\frac{2 e^2}{\pi h }\log(L/\ell_e),
\end{equation}
where $L$ is the system size (or coherence length) and $\ell_e$ is the mean free path. The negative sign implies a suppression of the conductivity leading to the weak localization (WL) regime, while a positive correction indicates that quantum interference constructively increases the diffusive conductivity, giving rise to WAL. Within the framework of this wavepacket propagation approach, one can then change the effective system size $L$ through its relationship with the mean square displacement $L \equiv \sqrt{\Delta X^2(\varepsilon,t)}$, and dynamically evaluate WL and WAL corrections by contrasting the quantum and the semi-classical conductivities. Additionally, this method also  allows for the calculation of the momentum relaxation time $\tau_\text{p} =\frac{\sigma_{\text{sc}}(\varepsilon_F)}{v_F^2\rho(\varepsilon_F)} $, a key parameter to characterize the scattering regimes\cite{Roche2012}.

The spin Hall conductivity is computed using a different real-space $\mathcal{O}(N)$ method based on the numerical evaluation of a modified version of the Kubo-Bastin formula \cite{Garcia2015, Garcia2016, VanTuan2016, Cresti2016}
\begin{align}
&\sigma_{{xy}}^{z}(\mu, T)=\frac{i\hbar}{\Omega}
\int_{-\infty}^{\infty}d\varepsilon f(T,\mu,\varepsilon)\label{DCconductivity}\\
&\times \text{Tr}\left\langle j_x\delta(\varepsilon-H)j_y^{z} \d{G^{+}(H,\varepsilon)}{\varepsilon}-j_x\d{G^{-}(H,\varepsilon)}{\varepsilon}j_y^{z} \delta(\varepsilon-H)\right \rangle , \nonumber
\end{align}
where $\delta(\varepsilon-H)$ is the $\delta$-function operator,  $j_y^z\equiv \left\{\sigma_z,j_y\right\}$ the spin-current operator with $\sigma_z$ the third Pauli matrix, and  $j_\alpha\equiv -i\frac{e}{\hbar}[H,r_\alpha]$ the $\alpha$-component of the current operator, $G^+(H,\varepsilon)$ and $G^-(H,\varepsilon)$ are the advanced and retarded Green's functions, and $f(T,\mu,\varepsilon)$ is the Fermi-Dirac distribution. In this method, the Green's functions and the $\delta$-functions are numerically calculated using the kernel polynomial method (KPM) \cite{Covaci2010,Garcia2015,Weisse2006,SILVER1994}.  For the determination of the spin Hall angle, defined as the charge to spin conversion efficiency $\gamma_{\text{sH}}= |\sigma_{\text{xy}}^{\text{z}}/\sigma_{\text{xx}}|$,  the diagonal conductivity in the denominator was also computed using the KPM method, in order to keep both within the same approximation.

Both methods allow us to simulate large system sizes; the quantum conductivity calculations in this article were performed on a $500\,\text{nm}^2$ system consisting of 9.2 million atoms, while for the spin Hall conductivity simulations we considered a $400\,\text{nm}^2$ system consisting of 8.3 million atoms.  The Chebyshev expansion of the Green's functions was performed with $M=4096$ expansion moments\cite{Weisse2006,SILVER1994} and the Jackson kernel, which is equivalent to a Gaussian broadening of $~10$ meV \cite{Jackson1912}. The trace in both cases was determined stochastically by using exponential random phase vectors \cite{Iitaka2004}. The spin Hall conductivities were computed within the PRACE infrastructure using 2000 cores per run, accumulating a total of 10 million hours.

\section{Results }

\begin{figure}[t]
%centering
\includegraphics[height=6cm]{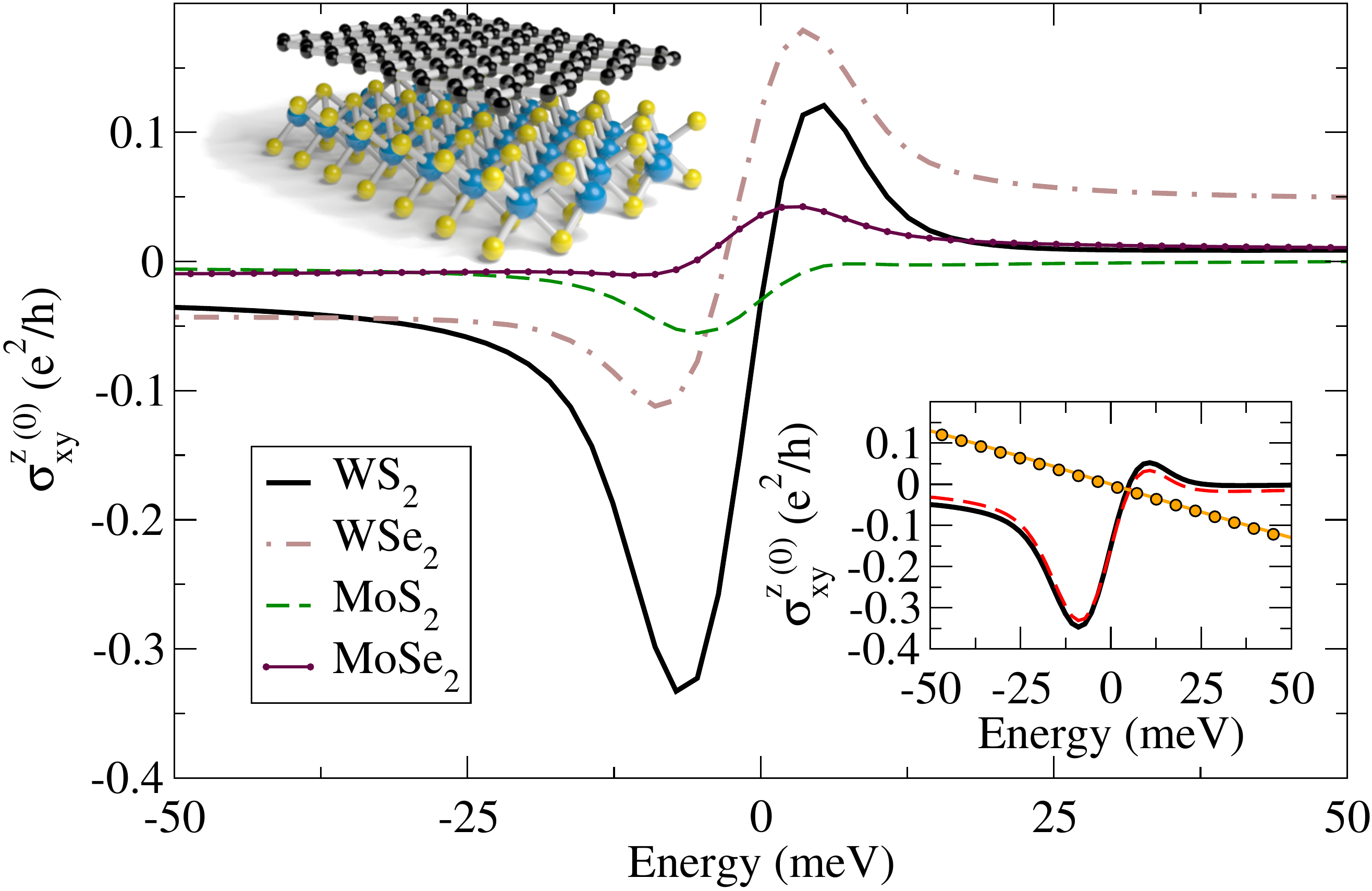}
\caption{Spin Hall conductivity $\sigma_{\text{xy}}^{\text{z}\,\,(0)}(\varepsilon_F)$ in the absence of disorder for different graphene/TMDC heterostructures, which serves as a proxy for the capability of a system to host the SHE. In the inset we show the spin Hall conductivity arising from only the sub-lattice-dependent intrinsic SOC (red and black curves, see text), indicating that this term is responsible for the peak in the main panel. We also show a comparison between the analytical calculation \cite{Dyrda2009} and our numerical approach for the pure Rashba case with $\lambda_\text{R}=0.1t$,  to validate our methodology (orange curve and symbols).}
\label{fig1}
\end{figure}

In Fig. \ref{fig1} we show the spin Hall conductivity for different graphene/TMDC heterostructures in the absence of disorder $\sigma_{\text{xy}}^{\text{z}\,\,(0)}(\varepsilon_F)$. In contrast to the dissipative conductivity, which is defined only in the presence of disorder, the spin Hall conductivity for a perfect crystal is proportional to the Berry phase \cite{Thouless1982}, which gives the topological contribution to spin transport \cite{Dyrda2009, RevModPhys.82.1959, EzawaHallEffect}. In the presence of a scalar disorder such as electron-hole puddles, this contribution is expected to be the main source of the intrinsic spin Hall effect, and therefore can be used as a reference to examine the potential for charge-spin conversion of a given material under these conditions. Fig. \ref{fig1} shows that all graphene/TMDC heterostructures possess non-zero spin Hall conductivity, which is consistent with previous experimental results \cite{Wang2016}. Additionally, we find that the graphene/WS$_2$ heterostructure stands out as the most promising for large spin Hall angles. In the inset of Fig. \ref{fig1}, we show $\sigma_{\text{xy}}^{\text{z}\,\,(0)}$ arising only from the sublattice-dependent intrinsic SOC. The magnitude of the peak near $\varepsilon_F = 0$ indicates that the main contribution to the spin Hall conductivity comes from this term, which can be separated into two contributions \cite{Yang2016}, a standard intrinsic SOC and a valley Zeeman SOC that couples spin and valley degrees of freedom. Due to this spin-valley coupling, one can conjecture that the relation between intra- and intervalley scattering can greatly impact the spin Hall effect in these heterostructures.

\begin{figure}[t]
\includegraphics[height=6cm]{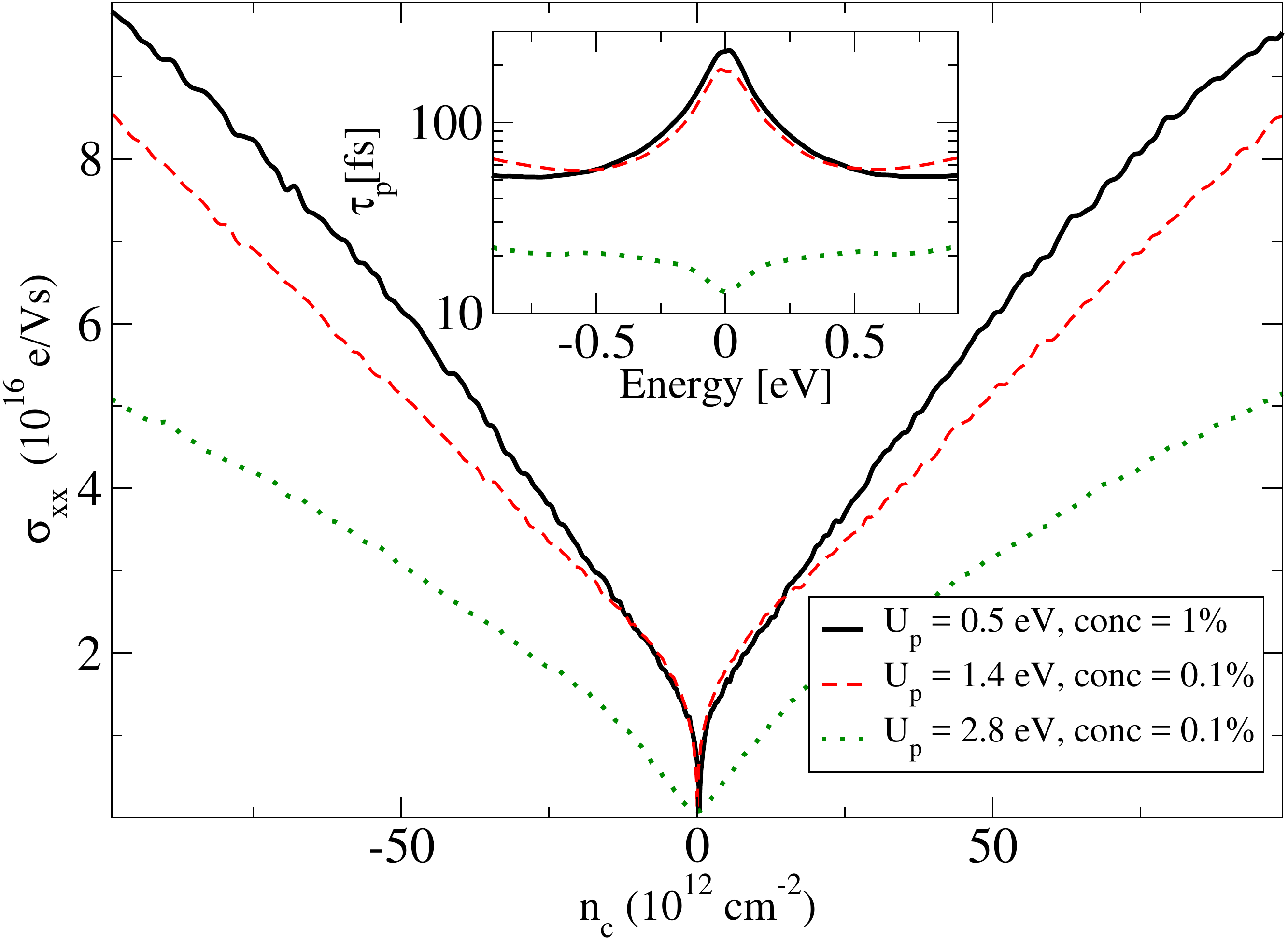}
\caption{ Diffusive conductivity for graphene/WS$_2$ heterostructures as a function of the carrier density for different kinds of disorder: $U_p=0.5$ eV and conc = $1\%$ (solid black line), $U_p=1.4$ eV and conc = $0.1\%$ (dashed red line), $U_p=2.8$ V and conc = $0.1\%$ (dotted green line). Inset: momentum relaxation time as a function of the Fermi energy. }
\label{fig2}
\end{figure}

Although $\sigma_{\text{xy}}^{\text{z}\,\,(0)}$ can be used as a starting point to determine the potential for spin Hall effects in these heterostructures, one also needs to consider the effect of disorder to correctly compute the spin Hall angle. For this purpose, we incorporate electron-hole puddle disorder into our system, and we follow the conclusions of previous studies \cite{Zhang2009, Roche2012} to control the ratio between the intra- and intervalley scattering rates $\pi_{\text{i.v}}\equiv\tau_{\text{intra}}^{-1}/\tau_{\text{inter}}^{-1}$. We consider two types of puddles with similar mobilities $\mu\approx 10^4~\text{cm}^2/\text{Vs}$ but different values of $\pi_{\text{i.v}}$: $U_p=0.5$ eV and conc = $1\%$, versus $U_p=1.4$ eV and conc = $0.1\%$, with ratio $\pi_{\text{i.v}}$ 1000 and 160 respectively  \cite{Zhang2009}. We also consider a third type of puddle with $U_p=2.8$ eV and conc = $0.1\%$, which has a lower mobility $\mu\approx 1000~\text{cm}^2/\text{Vs}$ and a much smaller value of $\pi_{\text{i.v}}=5\,$  \cite{Zhang2009}. The charge mobilities were obtained from the conductivity simulations, computed using the KPM method, presented in Fig. \ref{fig2},  in this plot we also show the momentum relaxation time $\tau_p$ as a function of Fermi energy (inset). The transition from a peak to a dip of $\tau_p$ at the Dirac point can be linked to the transition between pure intravalley scattering and a mix of inter- and intravalley scattering \cite{Roche2012}.

\begin{figure}[h]
%centering
\includegraphics[height=7cm]{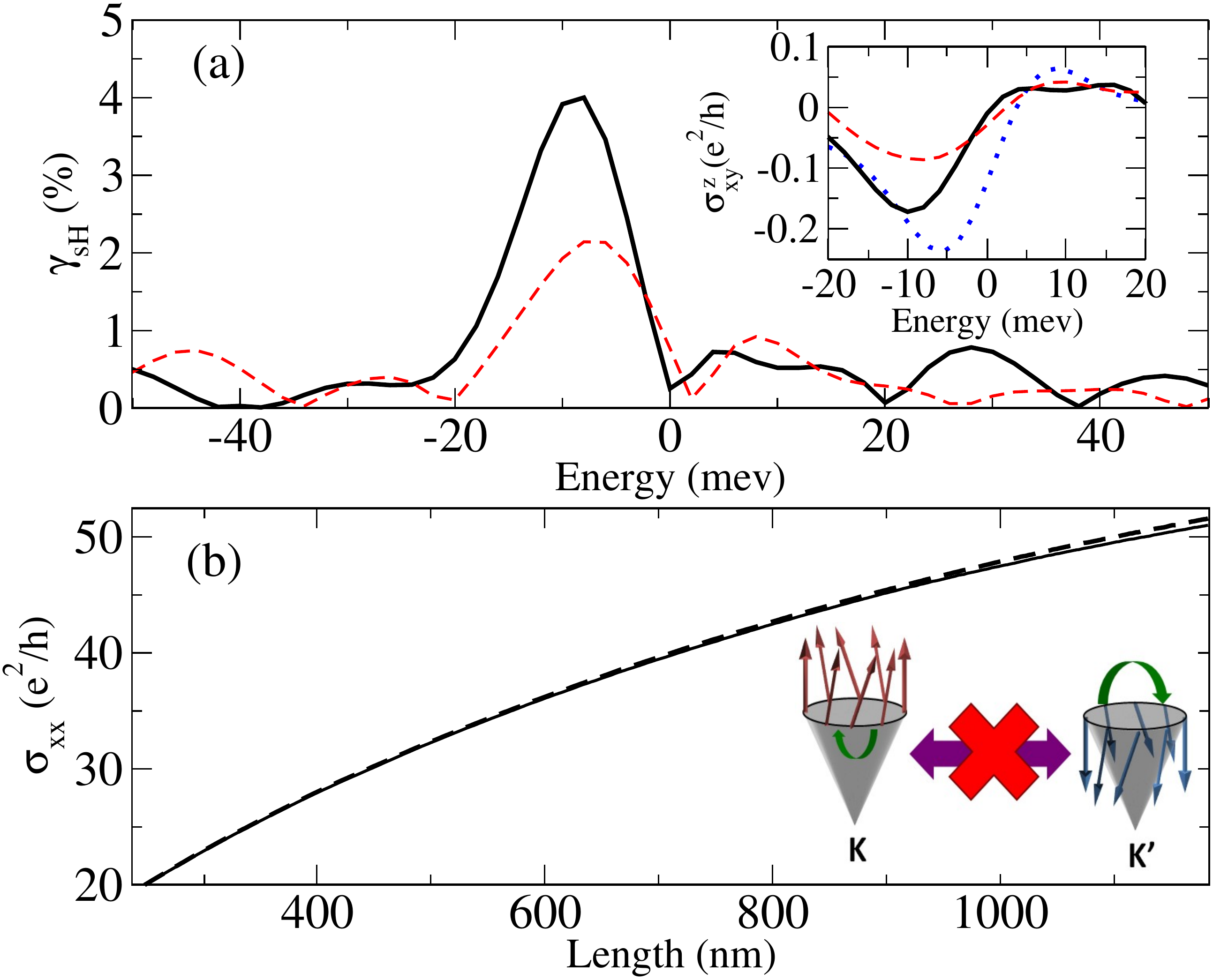}
\caption{ (a) Spin Hall angle $\gamma_{\text{sH}}$ for two puddle characteristics : $U_p=2.8$ eV and conc=$0.1\%$ (thick black line)  and $U_p=1.4$ eV and conc=$0.1\%$ (dashed red line). Inset: Spin Hall conductivity for the same puddles and in absence of disorder (dotted blue line) (b) Diffusive conductivity for graphene/$\text{WS}_2$ heterostructure at the Dirac point with (solid line) and without (dashed line) SOC for the puddle parameters $U_p=0.5$ eV, conc=$1\%$ . }
\label{fig3}
\end{figure}

\begin{figure}[h]
%centering
\includegraphics[height=7cm]{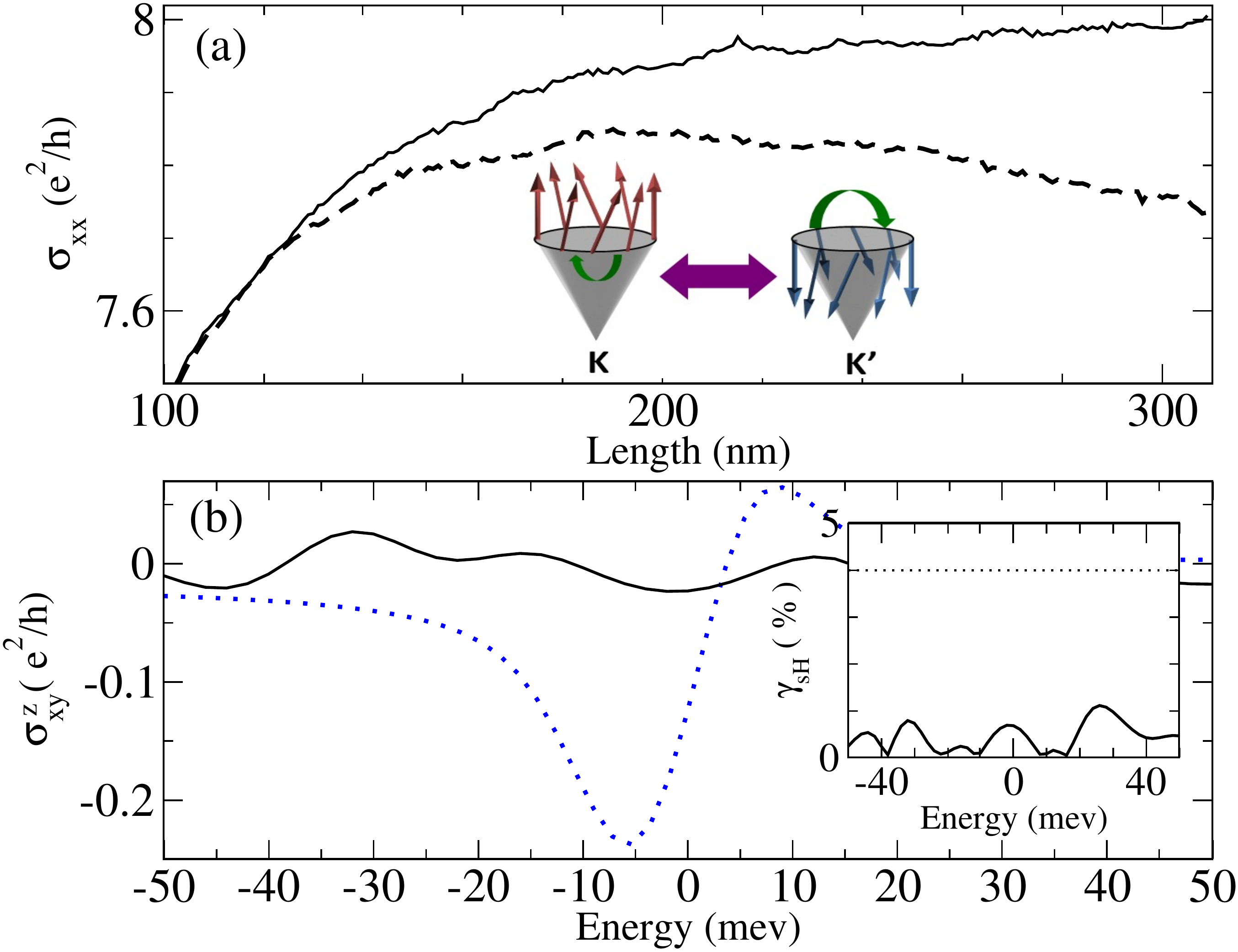}
\caption{ (a) Kubo conductivity for graphene/$\text{WS}_2$ at the Dirac point with (solid line) and without (dashed line) SOC for the puddle parameters $U_p=2.8$ eV, conc=$0.1\%$. (b) $\sigma_{\text{xy}}^{\text{z}}$ for the same puddle parameters (solid black line), with the clean case is shown for comparison (dotted blue line). Inset: Spin Hall angle for the same parameters; the maximum spin Hall angle obtained for the weak puddles is also shown as a reference (dotted line).}
\label{fig4}
\end{figure}

In Fig. \ref{fig3}(a) we show the spin Hall angle $\gamma_{\text{sH}}= |\sigma_{\text{xy}}^{\text{z}}/\sigma_{\text{xx}}|$ computed in the graphene/WS$_2$ system for these puddles. The highest charge-to-spin conversion was found to be $4\%$ for the shallowest puddles ($U_p=0.5$ eV). Interestingly, despite the similar mobilities between the shallow and intermediate ($U_p = 1.4$ eV) puddle disorder, the spin Hall angle suffered a reduction of $50\%$ for the latter case. Meanwhile, the spin Hall signal is markedly suppressed for the case of strong puddles ($U_p = 2.8$ eV), see Fig.\ref{fig4}(b)-inset. This trend is also evident in the spin Hall conductivity for these three systems, see Fig.\ref{fig3}(a)-inset.

As shown earlier, the main contribution to the spin Hall effect comes from the sublattice-dependent intrinsic SOC induced in the graphene by the TMDC (Fig. \ref{fig1} inset). Although broadening induced by intravalley scattering processes could suppress the SHE arising from intrinsic SOC, it should also affect the mobility of the sample in a similar fashion, leaving the spin Hall angle insensitive to the intravalley scattering rate. Instead, we attribute this reduction to intervalley scattering. The spin Hall effect for this type SOC is indeed very sensitive to the spin-valley locking induced by proximity effects.  Since the state of a carrier is completely defined by its momentum $\bm{p}$, its spin $s_z=\pm1$, and its valley $\tau_z=\pm1$ quantum numbers, by invoking Haldane's argument \cite{Haldane1988}, one can state that in presence of an intrinsic SOC $( \lambda_{\text{I}}^{(A)} = \lambda_{\text{I}}^{(B)}=\lambda_{\text{I}} )$, the effective mass of an electron with given spin and valley is $m(s_z,\tau_z)=s_z\tau_z\lambda_{\text{I}}$, which changes sign when switching spin or valley. Thus, when intervalley scattering is dominant the electron is continuously scattered from one valley to the other, which changes the sign of its effective mass. On average, this process effectively reduces the mass of the carrier, leading to a suppression of SHE. With valley Zeeman SOC $( \lambda_{\text{I}}^{(A)} \neq \lambda_{\text{I}}^{(B)} )$ the argument is similar; the sign of $s_z$ is opposite in opposite valleys, such that intervalley scattering will also lead to an average reduction of the effective mass.

%{\color{red}To verify the correlation between ...}

With respect to transport measurements, WL and WAL are quantum corrections of the semi-classical conductivity which are strongly related to the nature of disorder and symmetry breaking effects at play in the system. Both WL and WAL originate from the scattering of coherent electrons around a closed loop, which they traverse in clockwise and anticlockwise directions and interferes at the point of intercept. In the absence (presence) of SOC, the interference at the intercept is constructive (destructive), giving rise to the WL (WAL) phenomenon.  In graphene, the relative strength of intra- versus intervalley scattering is essential because it dictates which localization phenomenon dominates the low temperature transport \cite{McCann2006,Ortmann2011}. In the absence of SOC and intervalley scattering, WAL can be obtained due to the pseudospin degree of freedom, but WL dominates as soon as valley mixing is increased by short range disorder \cite{McCann2006,Ortmann2011}. The presence of SOC is then essential for generating WAL in the presence of intervalley scattering, although its observation in graphene also depends on the nature of symmetry breaking effects, as theoretically demonstrated by McCann and Falko \cite{McCann2012}.

Fig. \ref{fig3}.(b) shows the diffusive conductivity for the weaker puddles, although a similar picture is obtained for the intermediate ones. As can be observed, the diffusive conductivity is slightly suppressed when the SOC is turned on (solid line). This is due to the modification of the band structure, which becomes slightly more massive and therefore favors backscattering \cite{Zhang2009}. Fig. \ref{fig4}.(a) shows the dissipative conductivity for the strongest puddles with (solid line) and without SOC (dashed line). Here one can observe a clear transition from WL in the absence of SOC to WAL once the SOC is turned on.  This is consistent with the fact that intervalley scattering plays a fundamental role in determining which localization phenomenon dominates \cite{McCann2006,McCann2012}. Additionally, the presence of SOC is expected to lead to WAL as long as mirror symmetry is broken\cite{McCann2006,McCann2012}, which is the case for this system. However,  Fig. \ref{fig4}.(b) shows that under this conditions, the spin Hall conductivity is significantly suppressed, and the spin Hall angle becomes much less than $1\%$, which is again, a consequence of the fluctuating effective magnetic field that the electron feel due to intervalley scattering.

\section{Discussions}

In summary, we computed the spin Hall conductivities for different graphene/TMDC heterostructures, and found that all of them have sizable amplitude, as needed to observe the spin Hall effect, although the graphene/WS$_2$ heterostructure seems to be the most efficient one. By incorporating electron-hole puddles, we simulated samples with mobilities $\mu\approx 10000~\text{cm}^2/\text{Vs}$ and found a spin Hall angle which can reach $\gamma_{\text{sH}}\sim 4\%$. The origin of $\gamma_{\text{sH}}$ is related to the sublattice-dependent intrinsic SOC, induced by proximity with the TMDCs.  By varying the ratio of disorder-induced valley mixing, we showed that SHE in graphene/TMDC heterostructures is highly sensitive to inter-valley scattering and is more strongly suppressed by short-range disorder and large valley mixing. This information is key for the design of experiments aiming at observing SHE in such systems, since structural defects and grain boundaries \cite{Avsar2011,Isacsson2017}, which cause intervalley scattering and are usually found in large-area graphene materials, are clearly detrimental to strong SHE signals. We also reported on the opposite impact of SOC on the spin Hall effect and weak antilocalization phenomena \cite{Wang2015, Wang2016, Yang2016}, suggesting that the clear observation of WAL, despite providing unquestionable evidence of proximity-induced SOC, might simultaneously indicate valley mixing and vanishing of the SHE.

It should be noted that the calculated values for $\gamma_{\text{sH}}$ do not reach the experimental values $>10\%$ that have been reported to date \cite{Avsar2014}. Additionally, one observes that the spin-orbit coupling parameters extracted from WAL experiments \cite{Wang2015}, differ by almost one order of magnitude compared with those obtained from DFT \cite{Gmitra2015}. Different attempts to fit these parameters attribute this difference to a weak strain field that pulls graphene closer to the TMDC, hence increasing the proximity effect. Nevertheless, our results showed rather large $\gamma_{\text{sH}}$ even without considering this effect of SOC enhancement. One can thus anticipate that by "artificially" increasing the SOC-parameters used in the tight-binding model, larger $\gamma_{\text{sH}}$ could be easily obtained. Finally we also mention that our findings give an upper limit for the {\it intrinsic spin Hall effect} since the origin of the spin separation is entirely driven by spin splitting and spin textures of the bands. On the other hand, different types of defects could also lead a local rescaling of the spin-orbit interaction, and thus also generate an {\it extrinsic spin Hall effect} which would be superposed on the intrinsic SHE.

We finally note that our findings could also be of interest in the context of multifunctional spintronic/valleytronic systems based on TMD/graphene heterostructures, in the light of recent demonstrations of spin field effect transistors \cite{Hueso2016, Saroj2017} and hybrid spin valves based on optically-stimulated spin injection \cite{Luo2017}.

\begin{acknowledgement}

The authors acknowledge the Severo Ochoa Program (MINECO, Grant SEV-2013-0295), the Spanish Ministry of Economy and Competitiveness (MAT2012-33911), and Secretari\'a de Universidades e Investigaci\'on del Departamento de Econom\'ia y Conocimiento de la Generalidad de Catalu\~na. The authors also acknowledge the European Union Seventh Framework Program under  grant agreement 604391 (Graphene Flagship). This work has been performed thanks to the computational resources awarded from PRACE and the Barcelona Supercomputing Center (Mare Nostrum), under Project No. 2015133194. 

\end{acknowledgement}

\bibliography{achemsomanuscript}

\end{document}